\documentstyle[11pt,epsfig]{article}
\def\be{\begin{equation}}
\def\ee{\end{equation}}

\def\gsim{\mathrel{%
\rlap{\raise 0.511ex \hbox{$>$}}{\lower 0.511ex
\hbox{$\sim$}}}}

\def\lsim{\mathrel{
\rlap{\raise 0.511ex \hbox{$<$}}{\lower 0.511ex
\hbox{$\sim$}}}}

\begin{document}

\title{QUESTIONING THE EQUIVALENCE PRINCIPLE}
\date{ \ }
\author{Thibault DAMOUR\\
{\normalsize{\it Institut des Hautes \'Etudes Scientifiques, 91440 
Bures-sur-Yvette, France}}
}
\maketitle

\begin{abstract}
The Equivalence Principle (EP) is not one of the ``universal'' principles
of physics (like the Action Principle). It is a heuristic hypothesis
which was introduced by Einstein in 1907, and used by him to construct
his theory of General Relativity. In modern language, the (Einsteinian) 
 EP consists
in assuming that the only long-range field with gravitational-strength
couplings to matter is a massless spin-2 field. Modern unification theories,
and notably String Theory, suggest the existence of new fields 
(in particular, scalar fields: ``dilaton'' and ``moduli'') with 
gravitational-strength couplings. In most cases the couplings of these new
 fields ``violate'' the EP. If the field is long-ranged, these EP
 violations lead to many observable consequences (variation of ``constants'',
non-universality of free fall, relative drift of atomic clocks,...).
The best experimental probe of a possible violation of the EP is to compare
the free-fall acceleration of different materials.
\end{abstract}

\section{Introduction}\label{sec1}

Newton realized that it is remarkable that all bodies fall with
the same acceleration in an external gravitational field, because
this means that ``weight'' (the gravitational interaction) happens 
to be proportional to ``mass'' (the universal measure of inertia).
 However, it took Einstein to fully comprehend the
importance of this ``equivalence'' between weight (gravity) and 
mass (inertia). In 1907 \cite{E07} 
Einstein introduced what he called the
``hypothesis of complete physical equivalence'' between a
gravitational field and an accelerated system of reference. He
used this ``equivalence hypothesis'' ~\cite{E07,E11} as a
heuristic tool to construct a physically satisfactory
relativistic theory of gravitation. A posteriori, the Einsteinian Equivalence
Principle (EP) boils down to the assumption that the gravitational
interaction be entirely describable by a universal coupling of matter 
(leptons, quarks, gauge fields and Higgs fields) to 
the ``metric'' tensor $g_{\mu \, \nu}(x^{\lambda})$, 
 replacing everywhere in the matter Lagrangian the 
usual, kinematical, special relativistic (Minkowski) metric 
$\eta_{\mu \nu}$.
 In field theory language, this assumption is equivalent to requiring
that the only long-range field mediating the gravitational interaction be a 
massless spin-2 field. Seen in these terms,
we see that the EP is not one of the basic principles of Nature 
(like, say, the Action Principle, or the correlated Principle
of Conservation of Energy). It is a ``regional''principle which
restricts the description of one particular interaction. 
An experimental ``violation'' of the EP would not at all shake
the foundations of physics (nor would it mean that Einstein's theory
is basically ``wrong'').
 Such a violation might simply mean that the gravitational
interaction is more complex than previously assumed,
 and contains, in addition to the
basic Einsteinian spin-2 interaction, the effect of another long-range
field. [From this point of view, Einstein's theory would simply
appear as being incomplete.] Here, we shall focus on possible additional scalar 
fields, as suggested by string theory. Gravitational-strength vector fields 
would 
also lead to EP violations, though with a different phenomenology.

\section{Present experimental tests of the Equivalence Principle}\label{sec2}
The equivalence principle entails that electrically neutral 
test bodies follow geodesics of the universal spacetime metric
$g_{\mu \, \nu}(x^{\lambda})$, and that all the 
non-gravitational (dimensionless) coupling constants of matter 
(gauge couplings, CKM mixing angles, mass ratios,$\ldots$) are 
non-dynamical, i.e. take (at least at large distances) some fixed 
(vacuum expectation) values, independently of where and when, in 
spacetime, they are measured. Two of the best experimental tests 
of the equivalence principle are:

(i) tests of the universality of free fall, i.e. of the fact that 
all bodies fall with the same acceleration in an external 
gravitational field; and

(ii) tests of the ``constancy of the constants''.

Laboratory experiments (due notably, in our century, to 
E\"otv\"os, Dicke, Braginsky and Adelberger) have verified the 
universality of free fall to better than the $10^{-12}$ level. For instance, 
the fractional difference in free fall acceleration of Beryllium 
and Copper samples was found to be~\cite{Su94}
\be
\left( \frac{\Delta a}{a} \right)_{\rm Be \, Cu} = (-1.9 \pm
2.5) \times 10^{-12} \, . \label{eq:adel}
\ee
See also the work~\cite{Baessler99} which obtained a $\pm 5.6 \times 10^{-13}$
limit on the difference in free fall acceleration of specially constructed
(earth-core-like, and moon-mantle-like) test bodies.

The Lunar Laser Ranging experiment~\cite{LLR} has also verified that the Moon 
and the Earth fall with the same acceleration toward the Sun to better than one 
part in $10^{12}$
\be
\left( \frac{\Delta a}{a} \right)_{\rm Moon \, Earth} = (-3.2 \pm
4.6) \times 10^{-13} \, . \label{eq:llr}
\ee

A recent reanalysis of the Oklo phenomenon (a 
natural fission reactor which operated two billion years ago in 
Gabon, Africa) gave a very tight limit on a possible time 
variation of the fine-structure ``constant'', namely~\cite{DD96}
\be
-0.9 \times 10^{-7} < \frac{e_{\rm Oklo}^2 - e_{\rm now}^2}{e^2} 
< 1.2 \times 10^{-7} \, , \label{eq:oklo1}
\ee
\be
-6.7 \times 10^{-17} \, {\rm yr}^{-1} < \frac{d}{dt} \ {\rm
ln} \ e^2 < 5.0 \times 10^{-17} \, {\rm yr}^{-1} \, .
\label{eq:oklo2}
\ee

Direct laboratory limits on the time variation of the fine-structure constant
 $e^2$ are less stringent than Eq.(\ref{eq:oklo2}). For recent results,
 see Ref. \cite{Salomon00}. [ See also the claim \cite{Webb01} for a 
 cosmological change of $e^2$ of the order of one part in $10^5$.]
 
The tightness of the experimental limits (\ref{eq:adel})--(\ref{eq:oklo2}) might 
suggest 
to apply Occam's razor and to declare that the equivalence 
principle must be exactly enforced. However, the theoretical 
framework of modern unification theories, and notably string 
theory, suggest that the equivalence principle must be violated. 
Even more, the type of violation of the equivalence principle 
suggested by string theory is deeply woven into the basic fabric 
of this theory. Indeed, string theory is a very ambitious attempt 
at unifying all interactions within a consistent quantum 
framework. A deep consequence of string theory is that 
gravitational and gauge couplings are unified. In intuitive 
terms, while Einstein proposed a framework where geometry and 
gravitation were united as a dynamical field $g_{\mu \nu} 
(x)$, i.e. a soft structure influenced by the presence of matter, 
string theory extends this idea by proposing a framework where 
geometry, gravitation, gauge couplings, and gravitational 
couplings all become soft structures described by interrelated 
dynamical fields. A symbolic equation expressing this softened, 
unified structure is
\be
g_{\mu \nu} (x) \sim g^2 (x) \sim G(x) \, . \label{eq:ggg}
\ee
It is conceptually pleasing to note that string theory proposes 
to render dynamical the structures left rigid (or 
kinematical) by general relativity. Technically, 
Eq.~(\ref{eq:ggg}) refers to the fact that string theory (as well 
as Kaluza-Klein theories) predicts the existence, at a 
fundamental level, of scalar partners of Einstein's tensor field 
$g_{\mu \nu}$: the model-independent ``dilaton'' field $\Phi 
(x)$, and various ``moduli fields''. The dilaton field, notably, 
plays a crucial role in string theory in that it determines the 
basic ``string coupling constant'' $g_s = e^{\Phi (x)}$, which 
determines in turn the (unified) gauge and gravitational coupling 
constants $g \sim g_s$, $G \propto g_s^2$, as exemplified by the tree-level  
low-energy effective action
\be
L_{\rm eff} = e^{-2\Phi} \left[ \frac{R(g)}{\alpha'} + 
\frac{4}{\alpha'} \, (\nabla \Phi)^2 - \frac{1}{4} \, F_{\mu 
\nu}^2 - i \overline{\psi} \, D \psi - \ldots \right] \, . 
\label{eq:eff}
\ee

A softened structure of the type of Eq.~(\ref{eq:ggg}), embodied 
in the effective action (\ref{eq:eff}), implies a deep violation 
of Einstein's equivalence principle. Bodies of different nuclear 
compositions fall with different accelerations because, for 
instance, the part of the mass of nucleus $A$ linked to the 
Coulomb interaction of the protons depends on the space-variable 
fine-structure constant $e^2 (x)$ in a non-universal, 
composition-dependent manner. This raises the problem of the 
compatibility of the generic string prediction (\ref{eq:ggg}) 
with experimental tests of the equivalence principle, such as 
Eqs.~(\ref{eq:adel}), (\ref{eq:llr}) or (\ref{eq:oklo2}). It is 
often assumed that the softness (\ref{eq:ggg}) applies only at 
short distances, because the dilaton and moduli fields are likely 
to acquire a non zero mass after supersymmetry breaking. However, 
a mechanism has been proposed~\cite{DP} to reconcile
in a natural manner the existence of a {\it massless} dilaton (or 
moduli) field as a
fundamental partner of the graviton field $g_{\mu \nu}$ with
the current level of precision $(\sim 10^{-12})$ of
experimental tests of the equivalence principle. The
mechanism of~\cite{DP} (see also~\cite{DN} for
metrically-coupled scalars) assumes that string loop effects modify the 
effective action (\ref{eq:eff}) by replacing the various factors $e^{-2 \Phi}$ 
by more complicated functions of $\Phi$, e.g. $B_F(\Phi)=e^{-2 \Phi} + c_0 + c_1 
e^{2 \Phi} + \ldots $ Then, the very small couplings necessary
to ensure a near universality of free fall, $\Delta a/a <
10^{-12}$, are dynamically generated by the expansion of the
universe, and are compatible with couplings ``of order unity''
at a fundamental level. Refs. ~\cite{Antoniadis01,GPV01} discuss possible 
implementations of this 
mechanism in certain string models.

The aim of the present contribution is to emphasize the rich
phenomenological consequences of long-range dilaton-like fields, and 
to compare the probing power of various tests of the EP.
For addressing this question
we shall (following Refs. \cite{DP,jplclock,epclock})
 assume, as theoretical framework, the class of
effective field theories suggested by string theory.

For historical completeness, let us mention that the theoretical 
framework which has been most considered in the phenomenology of 
gravitation, i.e. the class of
``metric'' theories of gravity~\cite{W81}, which includes most
notably the ``Brans-Dicke''-type tensor-scalar theories,
appears, from a modern perspective, as being rather artificial. 
This is good news because
the phenomenology of ``non metric'' theories is richer and
offers new experimental possibilities. Historically,
the restricted class of ``metric'' theories was introduced in
1956 by Fierz~\cite{F56} to prevent, in an {\it ad hoc} way,
too violent a conflict between experimental tests of the
equivalence principle and the existence of a scalar
contribution to gravity as suggested by the theories of
Kaluza-Klein~\cite{KK} and Jordan~\cite{J}. Indeed, Fierz was
the first one to notice that a Kaluza-Klein scalar would
generically strongly violate the equivalence principle. He
then proposed to restrict artificially the couplings of the
scalar field to matter so as to satisfy the equivalence
principle. The restricted class of
equivalence-principle-preserving couplings introduced by Fierz
is now called ``metric'' couplings. Under the aegis of Dicke,
Nordtvedt, Thorne and Will a lot of attention has been given
to ``metric'' theories of gravity 
and notably to their
quasi-stationary-weak-field phenomenology (``PPN framework'',
see, e.g.,~\cite{W81}).
Note, however, that
Nordtvedt, Will, Haugan and others (for references see~\cite{W81}) studied 
conceivable phenomenological
consequences of generic ``non metric'' couplings, without
using a motivated field-theory
framework to describe such couplings.

For updated reviews of the experimental tests of gravity see \cite{PDG,W01}.

\section{Generic effective theory of a long-range dilaton}

Motivated by string theory, we follow Refs. \cite{DP,jplclock,epclock}
and consider the generic class of
theories containing a long-range dilaton-like scalar field
$\varphi$. The effective Lagrangian describing these theories
has the form (after a conformal transformation to the ``Einstein 
frame''):
\begin{eqnarray}
L_{\rm eff} &=& \frac{1}{4q} R(g_{\mu\nu}) - \frac{1}{2q} \
(\nabla \varphi)^2 - \frac{1}{4e^2 (\varphi)} \ (\nabla_{\mu}
A_{\nu} - \nabla_{\nu} A_{\mu})^2 \nonumber \\
&-& \sum_A \ \left[\overline{\psi}_A \,
\gamma^{\mu} (\nabla_{\mu} -iA_{\mu}) \psi_A + m_A (\varphi)
\, \overline{\psi}_A \psi_A \right] + \cdots \label{eq:01}
\end{eqnarray}
Here, $q\equiv 4\pi \, \overline G$ where $\overline G$ denotes
a bare Newton's constant, $A_{\mu}$ is the electromagnetic
field, and $\psi_A$ a Dirac field describing some fermionic
matter. At the low-energy, effective level (after the breaking
of $SU(2)$ and the confinement of colour), the coupling of the
dilaton $\varphi$ to matter is described by the
$\varphi$-dependence of the fine-structure ``constant'' $e^2
(\varphi)$ and of the various masses $m_A (\varphi)$. Here,
$A$ is a label to distinguish various particles. [A deeper
description would include more coupling functions, e.g.
describing the $\varphi$-dependences of the $U(1)_Y$,
$SU(2)_L$ and $SU(3)_c$ gauge coupling ``constants''.]

The strength of the coupling of the dilaton $\varphi$ to the
mass $m_A (\varphi)$ is given by the quantity
\be
\alpha_A \equiv \frac{\partial \ {\rm ln} \ m_A
(\varphi_0)}{\partial \ \varphi_0} \, , \label{eq:02}
\ee
where $\varphi_0$ denotes the ambient value of $\varphi (x)$
(vacuum expectation value of $\varphi (x)$ around the mass
$m_A$, as generated by external masses and cosmological
history). For instance, the usual PPN parameter $\gamma -1$
measuring the existence of a (scalar) deviation from the pure
tensor interaction of general relativity is given 
by~\cite{DEF},~\cite{DP}
\be
\gamma -1 = -2 \ \frac{\alpha_{\rm had}^2}{1+\alpha_{\rm had}^2}
\, , \label{eq:03}
\ee
where $\alpha_{\rm had}$ is the (approximately universal)
coupling (\ref{eq:02}) when $A$ denotes any (mainly) hadronic
object.

The Lagrangian (\ref{eq:01}) also predicts (as discussed 
in~\cite{DP}) a link between the coupling strength (\ref{eq:02})
and the violation of the universality of free fall:
\be
\frac{a_A -a_B}{\frac{1}{2} (a_A + a_B)} \simeq (\alpha_A
-\alpha_B) \alpha_E \sim -5\times 10^{-5} \, \alpha_{\rm
had}^2 \, . \label{eq:04}
\ee
Here, $A$ and $B$ denote two masses falling toward an external
mass $E$ (e.g. the Earth), and the numerical factor $-5 \times
10^{-5}$ corresponds to $A= {\rm Be}$ and $B= {\rm Cu}$. More precisely, 
dilaton-like models predict a specific type of composition 
dependence \cite{DP,london} for EP-violating effects. Namely,

\be
\left(\frac{\Delta  a}{a}\right )_{AB} = \hat {\delta}_A - \hat {\delta}_B\, ,
\label{eq:10a}
\ee
with
\be
\hat{\delta}_A = - (\gamma - 1) \left[ c_B \left(\frac{B}{\mu}\right )_A + c_D 
\left(\frac{D}{\mu}\right )_A + 0.943 \times 10^{-5} \left(\frac{E}{\mu}\right 
)_A  \right ]\, .
\label{eq:10b}
\ee
Here $\mu$ denotes the mass in atomic mass units, $B \equiv N+Z$ the baryon 
number, $D=N-Z$ the neutron excess and $E=Z(Z-1)/(N+Z)^{1/3}$ a quantity 
proportional to nuclear electrostatic energy. The third term on the 
right-hand-side of Eq. (\ref{eq:10b}) is expected to dominate the other two. Eq. 
(\ref{eq:10b}) gives a rationale for optimizing the choice of materials in free 
fall experiments 
(see Ref.~\cite{london} for a detailed discussion).

\medskip
 In addition to modifications of post-Newtonian gravity, such as
Eq. (\ref{eq:03}), and to violations of the universality of free
fall, Eq. (\ref{eq:04}), 
the Lagrangian (\ref{eq:01}) also predicts a host of other
effects linked to the spacetime variability of the coupling ``constants''
of physics. Some of these effects are, in principle, measurable 
by comparing the rates of high-precision clocks based on different
time-keepers.

To discuss the probing power of clock experiments, we need to introduce other 
coupling strengths, such as
\be
\alpha_{\rm EM} \equiv \frac{\partial \ {\rm ln} \ e^2
(\varphi_0)}{\partial \ \varphi_0} \, , \label{eq:07}
\ee
measuring the $\varphi$-variation of the electromagnetic (EM)
coupling constant\footnote{Note that we do not use the
traditional notation $\alpha$ for the fine-structure constant
$e^2 / 4\pi \hbar c$. We reserve the letter $\alpha$ for
denoting various dilaton-matter coupling strengths. Actually,
the latter coupling strengths are analogue to $e$ (rather than
to $e^2$), as witnessed by the fact that observable deviations
from Einsteinian predictions are proportional to products of
$\alpha$'s, such as $\alpha_A \alpha_E$, $\alpha_{\rm had}^2$,
etc$\ldots$}, and
\be
\alpha_A^{A^*} \equiv \frac{\partial \ {\rm ln} \ E_A^{A^*}
(\varphi_0)}{\partial \ \varphi_0} \, , \label{eq:08}
\ee
where $E_A^{A^*}$ is the energy difference between two atomic
energy levels.

In principle, the quantity $\alpha_A^{A^*}$ can be expressed
in terms of more fundamental quantities such as the ones
defined in Eqs. (\ref{eq:02}) and (\ref{eq:07}). For instance,
in an hyperfine transition
\be
E_A^{A^*} \propto (m_e \, e^4) \ g_I \ \frac{m_e}{m_p} \ e^4 \
F_{\rm rel} (Z e^2) \, , \label{eq:09}
\ee
so that
\be
\alpha_A^{A^*} \simeq 2 \, \alpha_e -\alpha_p + \alpha_{\rm EM}
\left( 4+\frac{d \ {\rm ln} \ F_{\rm rel}}{d \ {\rm ln} \ e^2}
\right) \, . \label{eq:10}
\ee
Here, the term $F_{\rm rel} (Z e^2)$ denotes the relativistic
(Casimir) correction factor~\cite{Casimir}. Moreover, in any
theory incorporating gauge unification one expects to have the
approximate link~\cite{DP}
\be
\alpha_A \simeq \left( 40.75 - {\rm ln} \ \frac{m_A}{1 \ {\rm
GeV}} \right) \ \alpha_{\rm EM} \, , \label{eq:11}
\ee
at least if $m_A$ is mainly hadronic.

We refer to Refs. \cite{jplclock,epclock} for a discussion of 
various clock experiments within the theoretical
framework introduced above. The most promising experiments
are the differential ``null'' clock experiments of the type
proposed by Will~\cite{W81} and first performed by Turneaure
et al.~\cite{T83}. For instance, if
(following the suggestion of~\cite{PTM}) one locally compares
two clocks based on hyperfine transitions in alkali atoms with
different atomic number $Z$, one expects to find a ratio of
frequencies 
\be
\frac{\nu_A^{A^*} ({\bf r})}{\nu_B^{B^*} ({\bf r})} \simeq
\frac{F_{\rm rel} (Z_A \, e^2 (\varphi_{\rm loc}))}{F_{\rm
rel} (Z_B \, e^2 (\varphi_{\rm loc}))} \, , \label{eq:14}
\ee
where the local, ambient value of the dilaton field
$\varphi_{\rm loc} = \varphi ({\bf r})$ might vary because of the (relative)
motion of external masses with respect to the clocks
(including the effect of the cosmological expansion). The
directly observable fractional variation of the ratio
(\ref{eq:14}) will consist of two factors:
\be
\delta \ {\rm ln} \ \frac{\nu_A^{A^*}}{\nu_B^{B^*}} = \left[
\frac{\partial \ {\rm ln} \ F_{\rm rel} (Z_A \, e^2)}{\partial
\ {\rm ln} \ e^2} - \frac{\partial \ {\rm ln} \ F_{\rm rel}
(Z_B \, e^2)}{\partial \ {\rm ln} \ e^2} \right] \times \delta
\ {\rm ln} \ e^2 \, . \label{eq:15}
\ee
The ``sensitivity'' factor in brackets, due to the
$Z$-dependence of the Casimir term, can be made of order 
unity~\cite{PTM}, while the fractional variation of the
fine-structure constant is expected in dilaton theories to be
of order~\cite{DP,jplclock,epclock}
\begin{eqnarray}
\delta \ {\rm ln} \ e^2 (t) &=& -\ 2.5 \times 10^{-2} \
\alpha_{\rm had}^2 \ U(t) \nonumber \\
&-& 4.7 \times 10^{-3} \ \kappa^{-1/2} ({\rm tan} \ \theta_0)
\ \alpha_{\rm had}^2 \ H_0 (t-t_0) \, . \label{eq:16}
\end{eqnarray}
Here, $U(t)$ is the value of the externally generated
gravitational potential at the location of the clocks, and
$H_0 \simeq 0.5 \times 10^{-10} \ {\rm yr}^{-1}$ is the Hubble
rate of expansion. [The factor $\kappa^{-1/2} \ {\rm tan} \
\theta_0$ is expected to be $\sim 1$.]

\section{Comparing the probing powers of various experimental tests}

We can now use the theoretical predictions given above to
compare the probing powers of various experimental tests of
relativistic gravity. 

Let us first compare post-Newtonian tests to (present) tests of
the universality of free fall. Solar-system measurements
of the PPN parameter $\gamma$, using VLBI measurements \cite{Eubanks},
constrains (via Eq.~(\ref{eq:03})) the dilaton-hadron coupling to $\alpha_{\rm 
had}^2 <
10^{-4}$. By contrast, the present tests of the universality of
free fall yields a much better limit. Namely, combining the
experimental limit Eq.~(\ref{eq:adel}) with the theoretical prediction
Eq.~(\ref{eq:04})
shows that the (mean hadronic) dilaton coupling strength is
already known to be smaller than:
\be
\alpha_{\rm had}^2 \lsim 10^{-7} \, . \label{eq:06}
\ee

If we now consider the constraints coming from the observed
lack of variability of the ``constants'', we find that the best
current constraint on the time
variation of the fine-structure ``constant'' (deduced from the
Oklo phenomenon), namely Eq.~(\ref{eq:oklo2}),
yields from Eq. (\ref{eq:16}) above, $\alpha_{\rm had}^2 \lsim
3 \times 10^{-4}$. 

Therefore, among present experimental results, the best constraint
on dilaton-like models comes from free fall experiments and constrains
the basic parameter $\alpha_{\rm had}^2$ to the $10^{- 7}$ level.

Turning our attention from {\it present} tests to possible {\it future}
tests, let us mention the level of $\alpha_{\rm had}^2$
that they will (hopefully) probe. 
The Stanford Gyro experiment (Gravity Probe B) will measure soon
(via a precise measurement of gravitational spin-orbit effects)
$\alpha_{\rm had}^2$ to the $10^{- 5}$ level. The high-precision 
astrometric mission GAIA should measure $\gamma$, and therefore $\alpha_{\rm 
had}^2$, to 
the $10^{- 7}$ level.
Let us now use
the (rough) theoretical prediction (\ref{eq:16})  to
compare quantitatively the probing power of clock experiments
to that of free fall tests. Let us
(optimistically) assume that clock stabilities of order
$\delta \nu / \nu \sim 10^{-17}$ (for the relevant time scale)
can be achieved. A differential {\it ground} experiment (using
the variation of the Sun's potential due to the Earth
eccentricity) would probe the level $\alpha_{\rm had}^2 \sim
3\times 10^{-6}$. A geocentric satellite differential
experiment could probe $\alpha_{\rm had}^2 \sim 5\times
10^{-7}$. These levels are interestingly low, but not as low 
as the present
equivalence-principle limit (\ref{eq:06}). To beat the level
(\ref{eq:06}) one needs to envisage an heliocentric
differential clock experiment (a few-solar-radii probe within
which two hyper-stable clocks are compared). Such an
experiment could, according to Eq. (\ref{eq:16}), reach the
level $\alpha_{\rm had}^2 \sim 10^{-9}$. 
 It is, however, to be noted that a much refined free fall test
of the equivalence principle such as  MICROSCOPE (respectively, STEP)
 aims at measuring $\Delta a/a \sim
10^{-15}$ (resp. $10^{-18}$), which corresponds to the level
 $\alpha_{\rm had}^2 \sim 10^{-11}$ (resp. 
$10^{-14}$), i.e. two (resp. five) orders of magnitude better than any
conceivable clock experiment.

\section{Conclusions}

In summary, the main points of the present contribution are:
\begin{enumerate}

\item[$\bullet$] String theory suggests the existence of new
gravitational-strength fields, notably scalar ones
(``dilaton'' or ``moduli''), whose couplings to matter violate
the equivalence principle. These fields can induce a
spacetime variability of the coupling constants of physics
(such as the fine-structure constant). 
\item[$\bullet$] The generic class of dilaton theories defined
above provides a well-defined theoretical framework in
which one can discuss the phenomenological consequences of the
existence of a (long-range) dilaton-like field. Such a theoretical
framework (together with some assumptions, e.g. about gauge
unification and the origin of mass hierarchy) allows one to
compare and contrast the probing powers of various experimental
tests of gravity. This comparison suggests that free fall experiments
are our best hope of probing a small, long-range violation of
the Equivalence Principle.
\item[$\bullet$] Let us finally note that, independently of any theoretical
prejudice, the recent (probable) discovery that gravity exhibits 
``repulsive'' effects on  cosmological scales~\cite{Perlmutter} provides
additional motivation for questioning  General Relativity on large scales. 
\end{enumerate}

\newpage
\begin{centerline}
{LE PRINCIPE D'\'EQUIVALENCE MIS EN QUESTION}
\end{centerline}

\medskip

\begin{centerline}
{Thibault DAMOUR}
\end{centerline}
\begin{centerline}
{{\it Institut des Hautes \'Etudes Scientifiques, 91440 
Bures-sur-Yvette, France}}
\end{centerline}

\medskip\noindent
\begin{abstract}
Le Principe d'\'Equivalence (PE) n'est pas un des 
principes universels de la physique, mais plut\^ot une hypoth\`ese heuristique 
qui restreint le contenu en champ de l'interaction gravitationnelle. La 
th\'eorie des cordes sugg\`ere l'existence de champs scalaires (notablement le 
dilaton) dont les couplages \`a la mati\`ere ``violent" le PE. Les exp\'eriences 
de chute libre apparaissent comme l'outil le plus pr\'ecis pour mettre en 
\'evidence une violation \'eventuelle (\`a longue port\'ee) du PE.
\end{abstract}

En 1907 Einstein introduisit ``l'hypoth\`ese de l'\'equivalence physique 
compl\`ete" entre la gravit\'e et l'inertie. Cette hypoth\`ese heuristique le 
conduisit \`a la construction de la th\'eorie de la Relativit\'e G\'en\'erale. 
En termes modernes, le ``Principe d'\'Equivalence" (PE) se r\'esume \`a imposer 
que le seul champ qui propage l'interaction gravitationnelle soit un champ de 
spin 2 \`a masse nulle.

\medskip
La th\'eorie de la Relativit\'e G\'en\'erale unifie g\'eom\'etrie et gravitation 
sous la forme du champ d'espace-temps $g_{\mu \nu}(x^{\lambda})$, c.\`a.d. d'une 
structure ``molle" qui est influenc\'ee par la pr\'esence de mati\`ere. En 
revanche, la Relativit\'e G\'en\'erale stipule que toutes les constantes de 
couplage de la physique sont ``rigides", c.\`a.d. fix\'ees a priori, et 
ind\'ependantes de la pr\'esence de mati\`ere. En revanche,  en th\'eorie des 
cordes, toutes les structures physiques (g\'eom\'etrie, gravitation, constantes 
de couplage) deviennent ``molles", c.\`a.d. d\'ecrites par des champs qui 
varient dans l'espace-temps. Cette variabilit\'e des ``constantes de couplage" 
implique de multiples ``violations" du PE : non-universalit\'e de la chute 
libre, d\'erive relative des horloges, etc...

\medskip
En utilisant, comme cadre th\'eorique, une classe de th\'eories d\'ecrivant les 
couplages g\'en\'eriques d'un champ scalaire du type dilatonique on peut 
d\'ecrire la ph\'enom\'enologie des violations possibles du PE en fonction d'un 
certain nombre de quantit\'es non dimensionn\'ees, $\alpha_A$ (mesurant le 
couplage du dilaton \`a la mati\`ere du type $A$). Ce cadre th\'eorique permet 
de comparer quantitativement l'``efficacit\'e" avec laquelle diverses 
exp\'eriences (tests de l'universalit\'e de la chute libre, tests des effets 
post-Newtoniens, comparaison d'horloges, ...) peuvent sonder des violations 
\'eventuelles du PE. Cette comparaison indique que les tests de l'universalit\'e 
de la chute libre (comme\linebreak MICROSCOPE ou STEP) sont notre meilleur 
espoir de 
d\'etecter une violation \'eventuelle du PE. Ind\'ependamment de toute 
th\'eorie, la r\'ecente d\'ecouverte (probable) d'effets gravitationnels 
``r\'epulsifs" \`a l'\'echelle cosmologique donne une motivation 
suppl\'ementaire pour mettre en question le comportement \`a longue port\'ee de 
la gravitation.

\end{document}